# Study On Universal Lossless Data Compression by using Context Dependence Multilevel Pattern Matching Grammar Transform

*Kim Chung Song,* Prof. Dr. *Kim Chol Hun*

College of Mathematics, **Kim Il Sung** University

**Abstract** In this paper, the context dependence multilevel pattern matching(in short CDMPM) grammar transform is proposed; based on this grammar transform, the universal lossless data compression algorithm, CDMPM code is then developed. Moreover, it is proved that this algorithms' worst case redundancy among all individual sequences of length n from a finite alphabet is upper bounded by $C(1/\log n)$ where $C$ is a constant.

**Key words** lossless data compression, MPM grammar transform

A grammar transform is a transformation that converts any data sequence to be compressed into a grammar from which the original data sequence can be fully reconstructed.

In grammar based-code, a data sequence is first converted into a grammar by a grammar transform and then losslessly encoded.

In paper [1], multilevel pattern matching grammar transform was proposed, based on this grammar transform, the lossless data compression code, MPM code were then developed. It is also proved that their worst case redundancies among all individual sequences of length n are upper bounded by $O(1/\log n)$. In paper [2],CMPM grammar transform that MPM grammar transform were extended to taken the side information was proposed, based on this grammar transform, CMPM code were developed and also proved that its worst case redundancy is upper bounded by $C(1/\log n)$ where C is constant.

In this paper, CDMPM grammar transform which is an extension of MPM grammar transform under the consideration of context, is proposed and based on this grammar transform, CDMPM code is then developed. It is also proved its worst case redundancies among all individual sequences of length n against the k-refinement context empirical entropy are upper bounded by $C(1/\log n)$ where C is a constant.

## 1. CDMPM Grammar Transform

Let $A = \{a_1, a_2, \cdots, a_{|A|}\}$ be the input data sequence alphabet.

For any positive integer n, $A^n$ denotes the set of all sequences of length n from $A$.
A sequence from A is sometimes called an *A*-sequence.

The input data sequence $x^n = x_1 x_2 \cdots x_n \in A^n$ is transformed into a context dependence grammar $G(x^n \mid C(x^n))$ or its equivalent form by CDMPM grammar transform.

Let $I \geq 1$ and $r \geq 2$ be two positive integers and their corresponding grammar transform is called the CDMPM$(r, I)$ grammar transform.

To get the transformed grammar $G(x^n | C(x^n))$ or its equivalent form given by the CDMPM$(r, I)$ grammar transform, we perform multilevel pattern matching.

First, we partition the string $x_1 x_2 \cdots x_n$ from left to right into non overlapping substrings of length $n_i$ $(i = I, I-1, \cdots, 0)$, where the lengths are obtained from the $r$-ary expansion of the integer n. we denote these substrings by $x^{(n_I)}, x^{(n_{I-1})}, \cdots, x^{(n_0)}$.

Let $(h \lfloor \log_r n \rfloor \cdots h_I \cdots h_0)_r$ be the $r$-ary expansion of $n$.

Then $n_I = (h \lfloor \log_r n \rfloor \cdots h_I 0 \cdots 0)_r$, $n_{I-1} = (h_{I-1} 0 \cdots 0)_r, \ldots, n_0 = (h_0)_r$.

The CDMPM$(r, I)$ grammar transform generates multilevel representation of $x^n$ according to the following steps.

Initially, for level $I$, we partition the substring $x^{(n_I)}$ from left to right into non overlapping sub-block sequence of length $r^I$ and denote these sub-block sequences $\widetilde{X}^{(I)} = \widetilde{X}^{(I)}_1 \widetilde{X}^{(I)}_2 \cdots \widetilde{X}^{(I)}_{n_I / r^I}$ and performs the following steps 3-5.

**Step 1** For each j such that $X_j^{(i+1)} = s$, partition the $A$-block sequence $\widetilde{X}_j^{(i+1)}$ into r sub block sequence of length $r^i$.

**Step 2** Partition the sub string $x^{(n_i)}$ into $A$-bock sequences of length $r^i$ and concatenate these A-block sequences to the A-block sequences of length $r^i$ constructed at step 1 above and denote these this new concatenated sequence of A-blocks at the level $i$ by $\widetilde{X}^{(i)} = \widetilde{X}^{(i)}_1 \widetilde{X}^{(i)}_2 \cdots \widetilde{X}^{(i)}_{|\widetilde{X}^{(i)}|}$.

**Step 3** Construct the context sequence $\widetilde{C}^{(i)} = \widetilde{C}^{(i)}_1 \widetilde{C}^{(i)}_2 \cdots \widetilde{C}^{(i)}_{|\widetilde{X}^{(i)}|}$ of sequence $\widetilde{X}^{(i)} = \widetilde{X}^{(i)}_1 \widetilde{X}^{(i)}_2 \cdots \widetilde{X}^{(i)}_{|\widetilde{X}^{(i)}|}$ as follows. The initial context $\widetilde{C}^{(i)}_1$ of $\widetilde{X}^{(i)}_1$ is fixed some $A$-sequence of length $r^i$ and context $\widetilde{C}^{(i)}_j$ $(j = 2, \cdots, |\widetilde{X}^{(i)}|)$ correspond A-sequence representing by $\widetilde{X}^{(i)}_j$.

**Step 4** Visit every A-block in the sequence $\widetilde{C}^{(i)}$ from left to right, and label all identical A-blocks with the same integers and all distinct $A$-blocks with distinct integers in increasing order, starting with 1. Denote each label corresponding to an $A$-block $\widetilde{C}_j^i$ by $C_j^i$.

Let $A_c^i$ be the set all distinct labels in $C^{(i)} = C^{(i)}_1 C^{(i)}_2 \cdots C^{(i)}_{|\widetilde{X}^{(i)}|}$.

For every distinct label $\gamma \in A_C^{(i)}$, let $\widetilde{X}^{(i)} | \gamma = \{\widetilde{X}_j^{(i)} : C_j^{(i)} = \gamma\}$. We call this sub sequence the context dependence subsequence corresponding to $\gamma$.

All context dependence subsequences of $\widetilde{X}^{(i)}$ are processed independently from each other at step 5 below.

**Step 5** For each distinct label $\gamma \in A_C^{(i)}$, visit every A-block in the context dependence subsequences $\widetilde{X}^{(i)} | \gamma$ from left to right and denote the first appearance of each distinct A-block in this subsequence by a special symbol '$s$'.

If the same $A$-block appears in $\widetilde{X}^{(i)} | \gamma$ again, label it by an integer so that all identical $A$-block in $\widetilde{X}^{(i)} | \gamma$, except for the most left one will be labeled by the same integer, which is just the number of distinct A-blocks in $\widetilde{X}^{(i)}|\gamma$ up to the first appearance of the A-block inclusively. For each A-block $\widetilde{X}_j^{(i)}$ in $\widetilde{X}^{(i)}$, denote its label by $X_j^{(i)}$ and $T_i = (X_1^{(i)} X_2^{(i)} \cdots X_{|T_i|}^{(i)})$.

For level 1, we perform only step 1,2 and 3 described above, and instead of performing step 4 and 5, let $T_0 = \widetilde{X}^{(0)}$, $C^{(0)} = \widetilde{C}^{(0)}$.

### 2. Arithmetic encoder

Input data sequence $x^n \in A^n$ is indirectly encoded by compressing the multilevel representation $T_i = (X_1^{(i)} X_2^{(i)} \cdots X_{|T_i|}^{(i)})$ $(i = I, \cdots, 0)$ of $X^n$.

Here the special symbol "s" assigned to the first A- block sequence in context $\gamma \in A_C^{(i)}$ dependence sequence is does not encoding. Before performing arithmetic coding, perform the following steps.

**Step 1** let $i = I$ and $L^{(I)} = T_I$.

**Step 2** let $L^{(i)} = l_1^{(i)} l_2^{(i)} \cdots l_{|L^{(i)}|}^{(i)}$.

Assume that $l_{j_1}^{(i)} = \cdots = l_{j_{\eta(s|L^{(i)})}}^{(i)} = s$ where $1 \leq j_1 < \cdots < j_{\eta(s|L^{(i)})}$ and $\eta(s|L^{(i)})$ is the total number of appearance of 's' in $L^{(i)}$. Replace each $l_{j_k}^{(i)}$ $(k = 1, \cdots, \eta(s|L^{(i)}))$ with the string $\hat{s} X_{kr-r+1}^{(i-1)} \cdots X_{kr}^{(i-1)}$ and denote the resulting sequence by $L^{(i-1)}$.

**Step 3** Repeat step 2 for $i = I-1, I-2, \cdots, 1$.

In the sequence $L^{(0)}$ obtained from step 3 above, replace each appearance of the symbol $\hat{s}$ with s, and remove all 's' corresponding to the first block in $\widetilde{X}^{(i)}|\gamma$, $\gamma \in A_C^{(i)}$, $i = I-1, I-2, \cdots, 1$ and denote the resulting sequence L.

Let $T_i'$ be the result sequence by removing the symbol "s" corresponding the first block of context dependence subsequence of $\gamma \in A_C^{(i)}$ in $T_i$ $(i = I, \cdots, 1)$. Here $T_0' = T_0$.

All symbols in sequence L is encoded independently according to which $T_i'$ it belong and what its context is.

Let $C_\gamma^i(\beta)$ be the counter for $\beta$ with context $\gamma \in A_C^{(i)}$ in $T_i'$.

Initially for $i = I, \cdots, 1$ $C_\gamma^i(\beta) = \begin{cases} 1, & \beta \in \{s, 1\} \\ 0, & \beta \notin \{s, 1\} \end{cases}$, and for $i = 0$, $C_\gamma^0(\beta) = 1$ $\gamma, \beta \in A$.

All symbols in $T_i'$ $(i = I, \cdots, 1)$ is encoded as follows.

**Step 1** encode $X_k^{(i)}$ by using the probability $C_\gamma^i(X_k^{(i)}) / \sum_\beta C_\gamma^i(\beta)$ where the summation $\sum_\beta$ is taken over $\{s, 1\} \cup \{1, \cdots, j\}$, and $j$ is the number of times that s occurs before the position of this 's' in $\widetilde{X}^{(i)}|\gamma$.

**Step 2** Increase the counter $C_\gamma^i(X_k^{(i)})$ by 1.

**Step 3** If $X_k^{(i)} = s$ then increase the counter $C_\gamma(j+1)$ from 0 to 1.

To encode $T_0'$, we use the following steps.

**Step 1** encode $X_k^{(0)}$ by probability $C_\gamma^0(X_k^0) \big/ \sum_{\beta \in A} C_\gamma^0(\beta)$ .

**Step 2** increase the counter $C_\gamma^0(X_k^0)$ by 1.

## 3. Compression rate related to the empirical entropy of the CDMPM grammar

Let $f_i^s$ be the number of symbol 's' in $T_i$ and $l_i = |T_i'|$.

Let $T_i'' = \{T_k''^{(i)}\} = \{X_j^{(i)} ; X_j^i \neq s\}$. Here $T_0'' = T_0$, $C_0'' = C_0$.

**Definition 1** we define the conditional empirical entropy of the CDMPM grammar $G(x^n | C(x^n))$ as follows.

$H_{G(x^n|C(x^n))} = \sum_{i=0}^{I} H(T_i'' | C_i'')$, where $H(T_i'' | C_i'')$ is the unnormalized conditional empirical entropy of the sequence $T_i''$ given $C_i''$.

That is $H(T_i'' | C_i'') = \sum_{\gamma \in A_{C_i}^{(i)}} \sum_{\beta \in \{1, 2, \cdots, f_i^s | \gamma\}} n(\beta) \log \dfrac{l_i'' | \gamma}{n(\beta)}$.

Here $n(\beta)$ is the number of occurrence of $\beta$ with context $\gamma$ in $T_i''$.

**Lemma** $\sum_{i=1}^{I} B_i$ be the size of the output binary codeword for the input sequence $x^n$. Then

$$\sum_{i=1}^{I} B_i \leq H_{G(x^n|C(x^n))} + 2\sum_{i=1}^{i} l_i - l_0 + |A|^2.$$

## 4. Redundancy of the CDMPM code

We now compare compression performance of the CDMPM algorithm with that of the best arithmetic coding algorithm with $k-$refinement context. Let $Z$ be a finite set consisting of $k$ elements; each $z \in Z$ is regarded as an abstract refinement context in addition to the contexts in $C$.

Let $P : (C \times Z) \times (A \times Z) \to [0, 1]$ be a transition probability function, satisfying

$$\sum_{a \in A, z \in Z} P(a, z | c, z') = 1, \quad c \in C, \quad z' \in Z.$$

Then for any sequence $x^n = x_1 x_2 \cdots x_n \in A^n$ the compression rate in bits per letter resulting from using the arithmetic coding algorithm with the transition probability $P$ to encode $x^n$ is given by $-\dfrac{1}{n} \sum_{z_1 \cdots z_n \in Z} \prod_{i=1}^{n} P(x_i, z_i | c_i, z_{i-1})$, where $z_0 \in Z$ is the initial refinement context.

Let $H_k(x^n | C(x^n)) = -\dfrac{1}{n} \log \left( \max_P \max_{z_0 \in Z} \sum_{z_1 \cdots z_n \in Z} \prod_{i=1}^{n} P(x_i, z_i | c_i, z_{i-1}) \right)$ and call the *k*-refinement-context empirical entropy of $x^n$. Where the maximization P varies over all probability function $P : (C \times Z) \times (A \times Z) \to [0, 1]$.

**Definition 2** we define the worst case redundancy of algorithm against the k-refinement empirical entropy $H_k(x^n | C(x^n))$ as $R_{n, k} = \max_{x^n \in A^n} \left[ \dfrac{1}{n} \sum_{i=0}^{I} B_i - H_k H_k(x^n | C(x^n)) \right]$.

**Theorem** $R_{n,k} < C/\log n$. Where $C = 20r \cdot \log|A| \cdot (r \cdot \log k + r \cdot \log|A| + |A|^2 + 2r + 1)$.

**Proof** fix $x^n = x_1 x_2 \cdots x_n \in A^n$ and $P$ be an transitional probability function which maximizes $H_k(x^n | C(x^n))$.

For any $x^q = x_1 x_2 \cdots x_q$, let $r(x^q | C(x^q)) = \max_{z_0 \in Z} \left( \sum_{z_1 \cdots z_n \in Z} \prod_{i=1}^{n} P(x_i, z_i | c_i, z_{i-1}) \right)$.

Then from definition of $r(x^q | C(x^q))$, it is fold $1 \leq \sum_{x^q \in A^q} r(x^q | C(x^q)) \leq k \cdot |A|$.

Also $X_1, X_2, \cdots, X_j$ be a nonoverlapping partition of $x^q$, it is hold that

$$r(x^q | C(x^q)) \leq r(X_1 | C(X_1)) \cdot r(X_2 | C(X_2)) \cdots r(X_j | C(X_j)).$$

For any $x^q \in A^q$, we normalize $r(x^q | C(x^q))$ over $A^q$ so that

$$P^*(x^q | C(x^q)) = \frac{Q_k}{k \cdot |A|} \cdot r(x^q | C(x^q))$$

$x^q \in A^q$ is a probability distribution satisfying $\sum_{x^q \in A^q} P^*(x^q | C(x^q)) = 1$. Then $1 \leq Q_k \leq k \cdot |A|$.

Hence $H_k(x^n | C(x^n)) = -\frac{1}{n} \log r(x^n | C(x^n)) \geq -\frac{1}{n} \sum\sum \log r(X_j'' | C_j''^{(i)})$ and

$$-\sum_{j=1}^{|T_i''|} \log r(X_j'' | C_j''^{(i)}) = -\sum_{j=1}^{|T_i''|} \log \left[ \frac{k \cdot |A|}{Q_k} \cdot P^*(X_j'' | C_j''^{(i)}) \right] \geq \sum_{j=1}^{|T_i''|} -\log P^*(X_j'' | C_j''^{(i)}) - |T_i''|(\log k + \log|A|).$$

From information theory $H(T_j'' | C_j'') = \min_{\widetilde{P}} \sum_{j=1}^{|T_i''|} -\log \widetilde{P}(X_j'' | C_j''^{(i)}) \leq \sum_{j=1}^{|T_i''|} -\log P^*(X_j'' | C_j''^{(i)})$

where the minimum is over all probability distributions $\widetilde{P}(\cdot | \cdot)$ on $A^q$

$$n \cdot H_k(x^n | C(x^n)) \geq \sum_{i=0}^{I} [H(T_j'' | C_j'') - |T_i''|(\log k + \log|A|)] \geq H_G(x^n | C(x^n)) - (\log k + \log|A|) \cdot \sum_{i=0}^{I} l_i$$

That is $\sum_{i=1}^{I} B_i - n H_k(x^n | C(x^n)) \leq 2\sum_{i=0}^{I} l_i + \sum_{i=1}^{I} f_i^s + (\log k + \log|A|) \cdot \sum_{i=0}^{I} l_i - l_0 + |A|^2$.

Meanwhile for large n, $\sum_{i=0}^{I} l_i \leq 20 r^2 \cdot \log|A| \cdot (n/\log n)$ and then

$$\frac{1}{n}\sum_{i=0}^{I} B_i - H_k(x^n | C(x^n)) \leq (\log k + \log|A| + 2 + 1/r) \cdot 20 r^2 \cdot \log|A| \cdot \frac{1}{\log n} + \frac{|A|^2}{n} \leq$$

$$\leq 20 r \cdot \log|A| \cdot (r \cdot \log k + r \cdot \log|A| + |A|^2 + 2r + 1)/\log n.$$

That is $R_{n,k} < C/\log n$, where $C = 20 r \cdot \log|A| \cdot (r \cdot \log k + r \cdot \log|A| + |A|^2 + 2r + 1)$. □